\documentclass{revtex4}
\usepackage{graphicx}
\usepackage{latexsym}
\usepackage{amsmath,amssymb}
\usepackage{epsfig,graphicx}
\def\be{\begin{equation}}
\def\ee{\end{equation}}
\def\bea{\begin{eqnarray}}
\def\eea{\end{eqnarray}}

\def\l{\left(}
\def\r{\right)}
\def\B{\Box_4}

\begin{document}


\title{Ghosts in the self-accelerating universe}
\author{Kazuya Koyama}
\email{kazuya.koyama.AT.port.ac.uk}%
\affiliation{Institute of Cosmology \& Gravitation, University of
Portsmouth, Portsmouth~PO1~2EG, UK}

\begin{abstract}
The self-accelerating universe realizes the accelerated 
expansion of the universe at late times by large-distance 
modification of general relativity without a cosmological constant. 
The Dvali-Gabadadze-Porrati (DGP) braneworld model provides an 
explicit example of the self-accelerating universe. 
Recently, the DGP model becomes very popular to study the 
observational consequences of the modified gravity models  
as an alternative to dark energy models in GR. 
However, it has been shown that the self-accelerating 
universe in the DGP model contains a ghost at the linearized level. 
The ghost carries negative energy densities and it leads to 
the instability of the spacetime. In this article, we review the origin of 
the ghost in the self-accelerating universe and explore the physical 
implication of the existence of the ghost.
\end{abstract}

\maketitle

\section{Introduction}
The acceleration of the late-time universe, as implied by
observations of Supernovae redshifts \cite{SN, SN2}, cosmic microwave background
anisotropies \cite{CMB} and the large-scale structure \cite{LSS}, 
poses one of the
deepest theoretical problems facing cosmology. Within the
framework of general relativity (GR), the acceleration must originate
from a dark energy field with effectively negative pressure, such
as vacuum energy or a slow-rolling scalar field (``quintessence").
So far, none of the available models has a natural explanation.
For example, in the simplest option of vacuum energy, leading to
the ``standard" LCDM model, the incredibly small,
\begin{equation}
\rho_{\Lambda,\mbox{\small{obs}}}={\Lambda \over 8\pi G}\sim H_0^2M_P^2
\ll \rho_{\Lambda,\mbox{\small{theory}}}\,,
\end{equation}
and incredibly fine-tuned,
\begin{equation}
\Omega_{\Lambda}\sim \Omega_m|_{\mbox{\small{today}}}\,,
\end{equation}
value of the cosmological constant cannot be explained by current
particle physics.

An alternative to dark energy plus GR is provided
by models where the acceleration is due to modifications of
gravity on very large scales, $r\gtrsim H_0^{-1}$ 
(see \cite{RM, KK0} and references therein). One of the
simplest covariant models is based on the Dvali-Gabadadze-Porrati
(DGP) brane-world model \cite{Dva}, in which gravity leaks off the
4D Minkowski brane into the 5D ``bulk"
Minkowski spacetime at large scales. The 5D action describing the DGP model 
is given by
\be
S = \frac{1}{2 \kappa^2}\int d^5 x \sqrt{-g} R 
+ \frac{1}{2 \kappa_4^2} \int d^4 x \sqrt{-\gamma} \:\: 
{}^{(4)\!}R - \int d^4 x \sqrt{-\gamma} {\cal L}_m.
\label{action0}
\ee
The important ingredient of the model is the 
induced Einstein-Hilbert term on the brane 
(see \cite{Aka0} for an early attempt). 
The existence of the brane imposes the junction 
condition for the metric at the position of the brane:
\begin{equation}
K_{\mu \nu} - K \eta_{\mu \nu} 
= -\frac{\kappa^2}{2} (T_{\mu \nu} - \kappa_4^{-2} G_{\mu \nu}),
\end{equation}
where $K_{\mu \nu}$ is the extrinsic curvature and 
we assume the refection ($Z_2$) symmetry across 
the brane. Due to the induced Einstein-Hilbert action,
the 4D Einstein tensor appears in the junction condition.

On small scales, gravity is
effectively bound to the brane and 4D Newtonian 
dynamics is recovered to a good approximation. The transition
from 4- to 5D behaviour is governed by a crossover 
scale $r_c$ \cite{Dva}
\be
r_c = \frac{\kappa^2}{2 \kappa_4^2}.
\ee 
The weak-field gravitational potential behaves as \cite{Dva}
\begin{equation}
\Psi \sim \left\{ \begin{array}{lll} r^{-1} & \mbox{for} & r< r_c,
\\ r^{-2} & \mbox{for} & r> r_c. \end{array}\right.
\end{equation}
The DGP model was generalized by Deffayet to a
Friedman-Robertson-Walker brane in a Minkowski bulk \cite{Deffayet:2000uy, Def2}; 
the gravity leakage at late times initiates acceleration -- not due to any
negative pressure field, but due to the weakening of gravity on
the brane. The energy conservation equation remains the same as in
GR, but the Friedman equation is modified:
\begin{eqnarray}
&& \dot\rho+3H(\rho+p)=0\,,\label{ec} \\ && H^2-{H \over r_c}=
{8\pi G \over 3}\rho\,. \label{f}
\label{Fried}
\end{eqnarray}

The modified Friedman equation~(\ref{f}) shows that at late times
in a CDM universe, with $\rho\propto a^{-3}\to0$, we have
\begin{equation}
H\to H_\infty= {1\over r_c}\,.
\end{equation}
Since $H_0>H_\infty$, in order to achieve acceleration at late
times, we require $r_c\gtrsim H_0^{-1}$, and this is confirmed by
fitting SN observations \cite{Deffayet:2002sp}.
Like the LCDM model, the DGP model is simple, with a single
parameter $r_c$ to control the late-time acceleration although the DGP
model does not provide a natural solution to the late-acceleration
problem; similarly to the LCDM model, where $\Lambda$ must be
fine-tuned, the DGP parameter $r_c$ must be fine-tuned to match
observation. 

\begin{figure}[h]
\centerline{
\includegraphics[width=13cm]{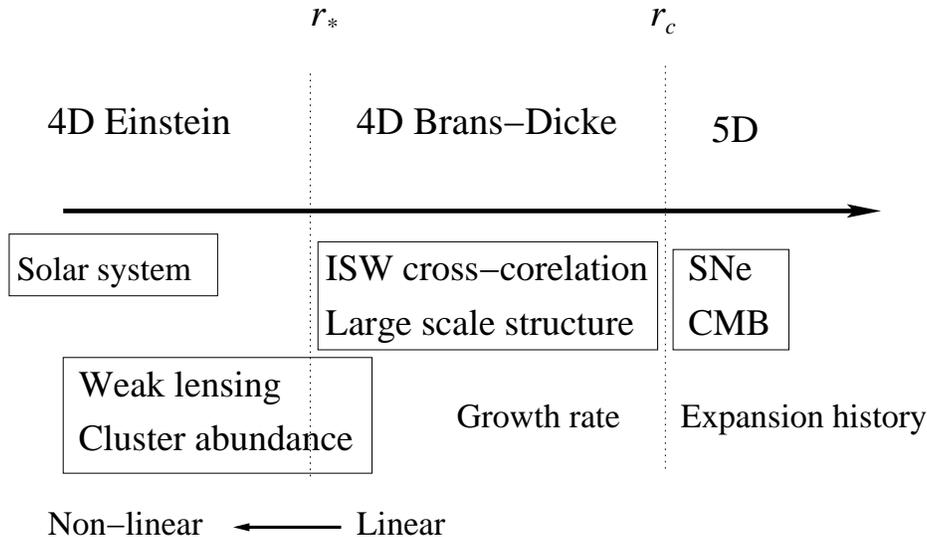}}
\caption{Summary of the behaviour of gravity in 
the DGP model. At large scales $r>r_c$, the theory 
is 5D. On small scales $r<r_c$, gravity becomes 4D but 
the linearized theory is described by a Brans-Dicke 
theory. This affects the large scale structure (LSS) and 
the Integrated Sachs-Wolfe (ISW) effect and its 
cross-correlation to LSS. Below the Vainstein radius
$r < r_*$, the theory approaches GR. This transition
can be probed by weak lensing and cluster abundance 
as the non-linear dynamics is important for these measures. 
The solar system tests also provide constraints on the model
in the 4D Einstein phase. From \cite{KK0}.
}
\label{fig1}
\end{figure}

The most interesting aspect of the DGP model is that there is a
possibility to distinguish the model from dark energy models in 
GR. This is because the recovery of the 4D 
GR on small scales is very subtle. 
Although the weak-field gravitational potential behaves as 4D on 
scales smaller than $r_c$, the linearized gravity is {\it not} described 
by GR. This is because there is no normalized 
zero-mode in this model and 4D gravity is recovered as 
a resonance of the massive KK gravitons. The massive graviton 
contains 5 degrees of freedom compared with 2 degrees of freedom 
in a massless graviton. One of them is a helicity-0 polarization. 
Due to this scalar degree of freedom, linearized gravity is described 
by Brans-Dicke (BD) gravity with vanishing BD parameter in the 
case of Minkowski spacetime. Thus this model would be excluded by 
solar system experiments. However, the non-linear interactions 
of the scalar mode becomes important on larger scales than expected \cite{Def3, 
Lue, Gru, Tan, LS1, Lue2, MS, Koy3}. 
Let us consider a static source with mass $M$. Gravity becomes non-linear 
near the Schwarzshild radius $r_g= 2 G M$. However, the scalar mode becomes 
non-linear at $r_* = (r_g r_c^2)^{1/3}$ (the Vainstein radius) 
which is much larger than $r_g$ if $r_c \sim H_0^{-1}$. In fact, for the Sun $r_*$ is much 
larger than the size of the solar system. 
A remarkable finding is at once the scalar mode becomes non-linear,
GR is recovered. This non-linear shielding of the 
scalar mode is crucial to escape from the tight solar system constraints.  
Fig.~\ref{fig1} summarizes the behaviour of gravity in the DGP model
(see \cite{LueR} for a review on the DGP model).

The fact that the linearized gravity is not described by 
GR offers an exciting possibility to 
distinguish this model from the dark energy models 
in GR. In cosmology, the modification
of linear perturbations affects the growth rate 
of the structure in the Universe. Thus structure formation 
in the DGP model is different from that in GR even if the 
expansion history of the Universe is exactly the same. 
Then combining the various observations, it is possible 
to distinguish the DGP model from the dark energy 
models in GR. Due to this possibility,
the DGP model has been a very popular model for modified gravity 
alternative to dark energy (for a review, see Ref.~\cite{KK0}). 

However, these interesting features of gravity could be signatures of 
a pathology of the model. The non-linear interaction of the scalar mode 
becomes important on very large scales compared with usual 4D 
gravity. If we consider a Planck mass particle, 
the Vainstein scale becomes $r_* =(\ell_{pl} r_c^2)^{1/3} \sim 1000$km where 
$\ell_{pl}$ is a Planck length and $r_c \sim H_0^{-1}$. This implies that 
quantum gravity corrections cannot be neglected below $1000$km 
\cite{Rubakov, LPR, NR}. 
Even if we focus on the linearized behaviour of the scalar mode, there 
appears a problem of a ghost instability
\cite{LPR, NR, KK1, KK2, Cha, Izu, Def5}. In the case where the brane is 
described by de Sitter spacetime, it has been proved that the scalar mode 
becomes a ghost. 

In this article, we focus on the problem of the ghost that appears 
in the self-accelerating universe. In section II, we study the 
spectrum of linearized perturbations about a de Sitter brane. 
Then we identify the origin of the ghost in section III. 
For the positive tension brane, the ghost is originated from 
a massive spin-2 graviton with mass $0 < m^2 < 2 H^2$. 
For the negative tension brane, the spin-0 perturbation 
associated from a fluctuation of the brane becomes a ghost. 
For a self-accelerating brane without tension, 
the ghost appears from the mixing between the spin-0 
and spin-2 perturbations.  
In section IV, the spectrum with matter source on the brane is 
studied. We highlight a difference between the DGP model 
and the massive gravity model that makes it difficult to remove 
the ghost by a simple modification of the model such as 
two-brane model. This ghost can be 
identified with the brane bending mode on small scales
which is a mix of helicity-0 components of massive spin-2 
perturbations and the spin-0 perturbation. 
In section V, the effective action for the brane bending mode is 
discussed. The leading order non-linear interaction 
is identified. It defines the Vainstein length below
which the linearized analysis cannot be trusted. 
We confirm again the existence of ghost at the linearized 
level. In section VI, fully non-linear solutions are 
discussed. These solutions indicate that the self-accelerating 
universe may suffer from instabilities even at 
non-perturbative level. Section VII is devoted to 
conclusions and discussions. 

\section{Perturbations about a de Sitter brane}
\subsection{Background solution}
Let us consider a situation where the brane is de Sitter 
spacetime. The bulk spacetime is a 5D Minkowski spacetime
and the metric is given by
\begin{equation}
ds^2 = dy^2 + N(y)^2 \gamma_{\mu \nu} dx^{\mu} dx^{\nu}, \quad
N(y)=1 \pm H |y|,
\end{equation}
where $\gamma_{\mu \nu}$ is the metric for the de Sitter spacetime
and the brane is located at $y=0$. The $Z_2$ symmetry across the 
brane is imposed. 
The junction condition at the brane gives the modified 
Friedmann equation
\begin{equation}
\pm \frac{H}{r_c} = H^2 - \frac{\kappa_4^2}{3} \sigma,
\end{equation}
where $\sigma$ is the tension of the brane. 
There are two branches of bulk solutions. The solution with 
$-$ sign is called the normal branch whereas the solution with 
$+$ sign is called the self-accelerating solution. These two 
different solutions correspond to different embeddings of the 
brane. The 4D de Sitter spacetime is a hyperboloid in a 5D 
Minkowski spacetime. In the self-accelerating universe, 
we take the outside of the hyperboloid as the bulk 
spacetime. On the other hand, in the normal branch, 
we take the inside of the hyperboloid. 

\begin{figure}[h]
\centerline{
\includegraphics[width=10cm]{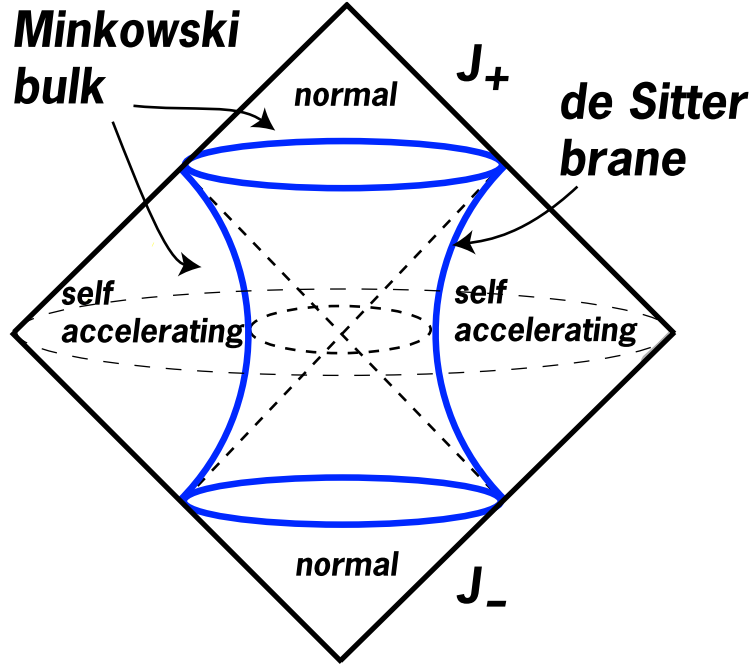}}
\caption{Embedding of a de Sitter brane in a
flat 5D bulk. The brane world volume is the hyperboloid in the
Minkowski bulk. The normal branch corresponds to
keeping the interior of the hyperboloid, and its mirror image
around the brane. In contrast, for the self-accelerating
branch, we keep the exterior, and its reflection. From \cite{Cha}.}
\end{figure}

\subsection{Spectrum of perturbations}
We study the linear perturbations 
\begin{equation}
ds^2 = dy^2 + (N(y)^2 \gamma_{\mu \nu} + h_{\mu \nu}) dx^{\mu} dx^{\nu},
\end{equation} 
about the background de Sitter spacetime. In addition to the 
gravitational perturbations $h_{\mu \nu}$, we must take into account a
perturbation of the position of the brane $y=\varphi(x)$ \cite{Tanaka}. 
Using the transverse-traceless gauge $\nabla^{\mu} h_{\mu \nu}=h=0$, 
the perturbed junction condition is given by \cite{KK1}
\begin{eqnarray}
k_{\mu \nu} &-& {\cal H} h_{\mu \nu} 
- r_c \left[ X_{\mu \nu}(h) 
-\kappa_4^2 \left(T_{\mu \nu} - \frac{1}{3} \gamma_{\mu \nu} T \right)
\right]\nonumber\\
&=& -(1-2 H r_c) 
\left(   
\nabla_{\mu} \nabla_{\nu} + H^2  \gamma_{\mu \nu} \right) \varphi,
\label{junction}
\end{eqnarray}
where ${\cal H} = N'/N|_{y=0} = \pm H$, $\nabla_{\mu}$ is a covariant derivative
with respect to $\gamma_{\mu \nu}$, 
$k_{\mu \nu} =(1/2) \partial_y h_{\mu \nu}$ on the brane is perturbations 
of the extrinsic curvature $K_{\mu \nu}$ and $X_{\mu \nu}$ is given by
\begin{eqnarray}
X_{\mu \nu} &=& \delta {}^{(4)} G_{\mu \nu} + 3 H^2 h_{\mu \nu} \nonumber\\
&=& -\frac{1}{2}\left(\B h_{\mu \nu} 
-\nabla_{\mu} \nabla_{\alpha} h^{\alpha}_{\nu} 
-\nabla_{\nu} \nabla_{\alpha} h^{\alpha}_{\mu} 
+ \nabla_{\mu} \nabla_{\nu} h \right) \nonumber\\
&-& \frac{1}{2} \gamma_{\mu \nu}(\nabla_{\alpha} \nabla_{\beta}
h^{\alpha \beta} - \B h)
+H^2 \left( h_{\mu \nu} +  \frac{1}{2} \gamma_{\mu \nu} h \right). 
\end{eqnarray}
The equation of motion for $\varphi$ is obtained from the traceless 
condition $h=0$;
\begin{equation}
(1 -2 {\cal H} r_c)(\B +4 H^2) \varphi 
= \frac{\kappa^2 T}{6}, 
\label{bending}
\end{equation}
where $\B =\nabla^{\mu} \nabla_{\mu}$.

Let us find solutions for the vacuum brane $T_{\mu \nu}=0$. 
Using the separation of variables $h_{\mu \nu} = \int dm~ e_{\mu \nu}(x) u_m(y)$, 
the equation of motion in the bulk is written as 
\begin{equation}
u_m'' + \frac{1}{N^2} (m^2 -2H^2) u_m=0, \quad (\B -m^2 -2H^2) e_{\mu \nu}(x)=0,
\end{equation}
where prime denotes a derivative with respect to $y$.
There are two types of solutions to Eq.~(\ref{junction}).
One type of the solution is a homogeneous solution with $\varphi=0$,
which is called the spin-2 perturbation. The spin-2 perturbations 
$\chi_{\mu \nu}$ satisfy the junction condition without $\varphi$
\begin{equation}
\chi_{\mu \nu}'-2 {\cal H} \chi_{\mu \nu} = -m^2 r_c \chi_{\mu \nu}.
\end{equation}
We find a tower of continuous Kaluza-Klein (KK) modes 
starting from $m^2 = (9/4) H^2$ as well as a normalizable discrete mode.
In the self-accelerating branch, the solution for the discrete mode
is given by 
\begin{equation}
u_{m_d} \propto N(y) ^{-1+1/Hr_c}, \quad \frac{m_d^2}{H^2} =\frac{1}{(Hr_c)^2}(3 Hr_c -1),
\label{solum}
\end{equation} 
for $Hr_c >2/3$ \cite{KK-0}. 
For $H r_c >1$, the mass is in the range $0 < m_d^2 \leq 2H^2$ where $m_d^2 = 2H^2$ for 
the self-accelerating universe $Hr_c=1$ and $m_d^2 \to 0$ for $H r_c \to \infty$. 
In the normal branch, the discrete mode is a zero mode $m_d^2=0$.

The other type of solution to Eq.~(\ref{junction}) is 
an inhomogeneous solution sourced by the scalar mode $\varphi$. 
We call this solution the spin-0 perturbation. 
In the self-accelerating branch, there is
a normalizable solution given by \cite{KK1}
\begin{equation}
h_{\mu \nu} = \frac{1-2Hr_c}{H(1-Hr_c)} (\nabla_{\mu} \nabla_{\nu} 
+ H^2 \gamma_{\mu \nu}) \varphi.
\label{solphi1}
\end{equation}
This is a solution with $m^2 = 2H^2$. In the normal branch, 
there is no normalisable spin-0 perturbation. Then, 
the normal branch is ghost-free. 
Eq.~(\ref{solphi1}) is singular at $Hr_c=1$ and we will 
deal with this case separately in section III.C.

\section{Ghost in de Sitter spacetime}
\subsection{Effective action}
We can construct the 2nd order action for $h_{\mu \nu}$ and $\varphi$
from the 5D action (\ref{action0}). The result is given by \cite{KK1}
\begin{equation}
\delta_2 S = -\frac{1}{4 \kappa^2} \int d^5 x \sqrt{-g} N^{-4} h^{\mu \nu} 
\delta {}^{(5)} G_{\mu \nu} 
+\frac{1}{\kappa^2} \int d^4 x\sqrt{-\gamma} {\cal L}_B,
\label{5Dactionp1}
\end{equation}
where $\delta {}^{(5)} G_{\mu \nu}$ is the 5D perturbed Einstein 
tensor and 
\begin{eqnarray}
{\cal L}_B &=& k^{\mu \nu} h_{\mu \nu} -k h
+ \frac{1}{2} {\cal H} (h^2 - h^{\mu \nu} h_{\mu \nu})\nonumber\\ 
&+& (1-2 {\cal H} r_c) 
\left(h_{\mu \nu} \nabla^{\mu} \nabla^{\nu} \varphi - h 
\nabla^{\rho} \nabla_{\rho} \varphi -3 H^2 h \varphi \right)
\nonumber\\
&-& 3{\cal H} \left(-(1-2 {\cal H} r_c)\varphi (\B + 4H^2) \varphi 
+ \frac{\kappa^2}{3} T \varphi \right) \nonumber\\
&+& \frac{1}{2} \kappa^2 h^{\mu \nu} T_{\mu \nu}  
-\frac{r_c}{2} h^{\mu \nu} X_{\mu \nu}(h).
\label{5Dactionp2}
\end{eqnarray}
This action gives the correct equation of motion and 
the junction condition for $h_{\mu \nu}$ and the equation of 
motion for $\varphi$. 

In the following discussions, we focus on the self-accelerating 
branch. We can derive an effective action for the
brane fluctuation $\varphi$ by substituting the 5D solution for $h_{\mu \nu}$
given by $\varphi$ (\ref{solphi1}) into the 5D action (\ref{5Dactionp1}) 
and get the off-shell 
action for $\varphi$ by integrating out only with respect to 
the extra coordinate $y$ \cite{Padilla}.
This yields the action for $\varphi$ as
\begin{equation}
S_{\varphi} = \frac{3 H}{2 \kappa^2} 
\left(\frac{1- 2Hr_c}{1- H r_c}\right) 
\int d^4 x \sqrt{-\gamma} \varphi (\B + 4 H^2) \varphi.
\label{actphi}
\end{equation}
The 4D effective action for the spin-2 perturbations is also obtained
in a similar way. For the discrete mode with $m_d^2$, we get
\begin{equation}
S_{\chi} = \frac{r_c (3 Hr_c-1)}{4 \kappa^2 (3 Hr_c-2)} 
\int d^4x \sqrt{-\gamma} \chi^{\mu \nu} (\B -2H^2 - m_d^2) \chi_{\mu \nu},
\label{actchi}
\end{equation}
where transverse-traceless gauge fixing conditions 
$\nabla^{\mu} \chi_{\mu \nu} =\chi^{\mu}_{\mu}=0$ are imposed . 
This is exactly the same action for the 
spin-2 perturbations in the 4D massive gravity theory 
where the Pauli-Fierz (PF) mass term is added to the Einstein-Hilbert action 
by hand \cite{PF} 
\begin{equation}
S_M = -\frac{M^2}{8 \kappa_4^2}\int d^4 x \sqrt{-\gamma}(h^{\mu \nu} h_{\mu \nu} - h^2).
\end{equation}

\subsection{Ghost in de Sitter spacetime with a tension}
Ref.~\cite{KK1} studied the existence of the ghost based on the 
above effective action. In the limit $Hr_c \to \infty$, the action 
(\ref{actchi}) approaches the one for massless spin-2 perturbations. 
However, there is a 
discontinuity between the massless perturbations and the massive 
perturbations that is known as the van Dam-Veltman-Zakharov 
discontinuity \cite{vDVZ}. 
Due to the lack of gauge symmetry, the massive
spin-2 perturbations contain a helicity-0 excitation. Moreover, 
it has been shown that this helicity-0 excitation becomes a ghost
if $0 < M^2 < 2H^2$ \cite{Higuchi,DW} (see also \cite{Cha}). 
This is exactly the same mass range for the 
discrete mode in the self-accelerating branch 
for $Hr_c > 1$. 
Thus we identify the ghost as the helicity-0 mode of 
the discrete mode of the spin-2 perturbations. 
On the other hand, for $Hr_c <1$, the spin-2 perturbations become 
healthy as the mass for the spin-2 perturbations is larger than 
$2H^2$. However, in this case, the coefficient in front of the 
effective action for the spin-0 perturbations becomes negative
and the spin-0 perturbation becomes a ghost. 

\begin{figure}[h]
\centerline{
\includegraphics[width=13cm]{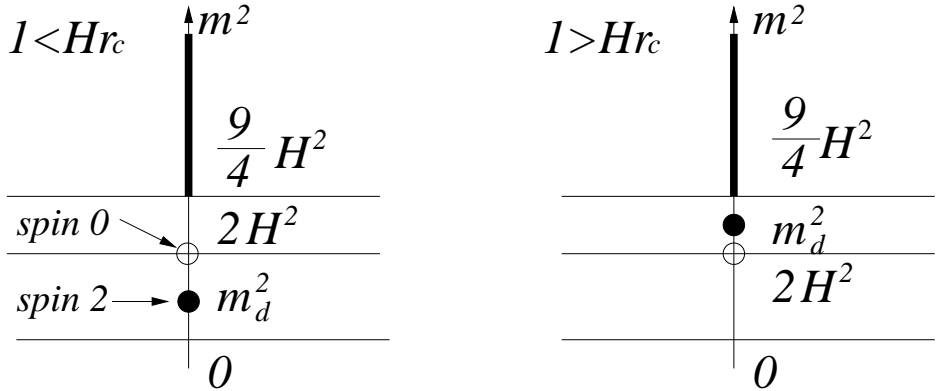}}
\caption{Summary of the mass spectrum of the
scalar perturbations in $+$ branch. 
Spin-2 perturbation has continuous modes with $m^2 \geq (9/4)H^2$
and a discrete mode with $m^2 = m_d^2$ while spin-0 perturbation
has $m^2 = 2H^2$.
In the limit
$Hr_c \to 1$, both the helicity-0 excitation
of spin-2 perturbation and the spin-0 perturbation have mass
$m^2 = 2H^2$ and there is a resonance.
}
\label{fig3}
\end{figure}

\begin{figure}[h]
\centerline{
\includegraphics[width=13cm]{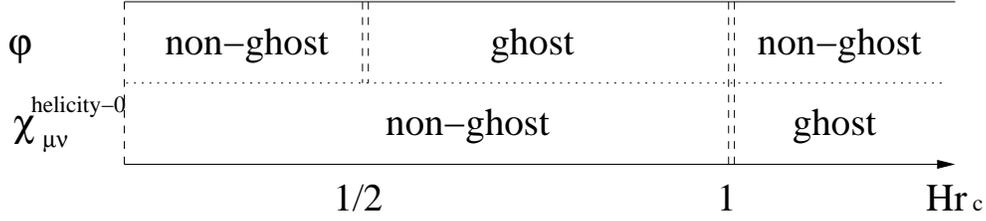}}
\caption{Summary of the existence of the ghost. From \cite{KK1}.
}
\label{fig4}
\end{figure}

\subsection{Ghost in self-accelerating universe}
In the self-accelerating universe $Hr_c=1$, the 
mass of the discrete mode of the spin-2 perturbations 
becomes $2H^2$. This is a special mass in the massive 
gravity as the action is invariant under the 
transformation
\begin{equation}
\chi_{\mu \nu} \to \chi_{\mu \nu} + (\nabla_{\mu} \nabla_{\nu} - H^2 \gamma_{\mu \nu})
X,
\label{enhance}
\end{equation}
where $X$ is any solution of the equation $(\B + 4H^2) X =0$.
This is so called enhanced symmetry and the helicity-0 mode
can be eliminated by this symmetry \cite{DW}. Then there are 
only four polaization when the mass is $m^2= 2 H^2$ and the ghost 
disappears. 

However, this does not happen in the self-accelerating 
universe as there is an additional spin-0 perturbations 
with the same mass \cite{KK2}. In fact in the self-accelerating universe, 
$Hr_c = 1$, the bulk wave function for the discrete mode 
is given by $u_{m_d} \propto N(y)$ and it is the same as the spin-0 
perturbation. Then the discrete mode of the spin-2 perturbations
and the spin-0 perturbation degenerate and they can mix. 
This comes from the fact that at $m^2=2H^2$,
the spin-0 and spin-2 perturbations degenerate. 
In fact we can make the spin-2 perturbations from a scalar
\be
h_{\mu \nu}^{(2H^2)} = (\nabla_{\mu} \nabla_{\nu} - H^2 \gamma_{\mu \nu}) X,
\quad (\B + 4H^2)X =0.
\label{gauge}
\ee
This is a scalar mode with mass squared $-4H^2$. At the same time 
$h_{\mu \nu}$ is a transverse-traceless perturbations and from the identity, 
\be
(\B - 4H^2) h_{\mu \nu}^{(2H^2)} \nonumber\\
=\left( \nabla_{\mu} \nabla_{\nu} - H^2 \gamma_{\mu \nu} \right) 
(\B+ 4H^2) X =0,
\ee
it is identified to have a mass given by $m^2 = 2 H^2$.

Let us investigate the limit $Hr_c \to 1$
carefully. The solutions for the metric perturbations 
are given by
\begin{equation}
h_{\mu \nu} = \chi^{m_d}_{\mu \nu}(x) u_{m_d}(y)
+ \frac{1-2H r_c}{H(1-Hr_c)} (\nabla_{\mu} \nabla_{\nu}+H^2
\gamma_{\mu \nu}) \varphi,
\end{equation}
where $u_{m_d}$ is given by Eq.~(\ref{solum}) and we only consider the localized 
modes. The limit $Hr_c \to 1$
looks singular, but we can perform a field re-definition 
\begin{equation}
\chi^{m_d}_{\mu \nu}(x)=A_{\mu \nu}
-\frac{1-2H r_c}{H(1-Hr_c)} (\nabla_{\mu} \nabla_{\nu}+H^2
\gamma_{\mu \nu}) \varphi.
\end{equation}
The metric perturbations read
\begin{equation}
h_{\mu \nu}(x,y)
=A_{\mu \nu} u_{m_d}(y)
+\frac{1-2H r_c}{H(1-Hr_c)} (\nabla_{\mu} \nabla_{\nu}+H^2
\gamma_{\mu \nu}) \varphi (1-u_{m_d}(y)).
\end{equation}
Then taking the limit $Hr_c \to 1$, we get \cite{KK2}
\be
\label{hAB}
h_{\mu\nu}=A_{\mu\nu}(x)
+\frac{1}{H}(\nabla_\mu\nabla_\nu+H^2\gamma_{\mu\nu})\varphi(x)
\log \l 1+Hy\r\;. 
\ee
Substituting this expressions into the bulk equation and 
the junction condition, we get an equation for $A_{\mu \nu}$
\be
\label{AB}
\B A_{\mu\nu}-4H^2 A_{\mu\nu}=H\l \nabla_\mu\nabla_\nu
+H^2\gamma_{\mu\nu}\r\varphi.  
\ee
 
The effective action for $A_{\mu \nu}$ and $\varphi$ 
is obtained by substituting the solutions for $A_{\mu \nu}$
and $\varphi$ into the 5D action and integrating out the 
extra-dimension as  
\be
\label{effact}
\begin{split}
S_{eff}=\frac{1}{\kappa^2H}\int d^4x\sqrt{-\gamma}
\bigg\{&-A^{\mu\nu}X_{\mu\nu}(A)-H^2A^{\mu\nu}A_{\mu\nu}
+H^2A^2\\
&-H(A^{\mu\nu}\nabla_\mu\nabla_\nu\varphi-A\B\varphi-3H^2A\varphi)
-\frac{9H^2}{4}\varphi(\B+4H^2)\varphi\bigg\}\;,
\end{split}
\ee
where we introduced the notation $A\equiv A_\mu^\mu$. 
The first line in Eq.~(\ref{effact}) coincides with 
the quadratic Lagrangian for the Pauli-Fierz theory 
of massive gravity with $m^2 = 2H^2$. Thus the action 
could have the enhanced symmetry (\ref{enhance}) 
if there were no mixing to 
$\varphi$. However, the mixing between $\varphi$ and 
$A_{\nu \mu}$ breaks this symmetry explicitly.

In order to study the existence of the ghost, it is necessary to 
write the effective action only in terms of physical degrees of 
freedom. Ref.~\cite{KK2} performed the Hamiltonian analysis to derive 
the reduced Hamiltonian written only in terms of physical 
degrees of freedom. There are two dynamical degrees of freedom,
namely the spin-0 mode and the helicity-0 excitation of spin-2
perturbations. It is found that, in general, the Hamiltonian 
cannot be diagonalized. However, on small scales under horizon,
it is possible to diagonalize the Hamiltonian and we find a ghost
from a mixing between the spin-0 perturbation and the helicity-0
excitation of spin-2 perturbations.

\subsection{Two-brane model}
Ref.~\cite{Izu} tried to remove the ghost 
by adding the second brane in the bulk. 
In previous section, we saw that the origin of the 
ghost for the positive tension brane is coming 
from the fact the discrete mode for the 
spin-2 perturbations has a mass in the range 
$0 < m^2 < 2 H^2$. 
It is possible to eliminate this ghost by 
introducing the second brane in the bulk. 
By making the distance between the two branes 
short, the mass increases as in the usual 
Kalzua-Klein compactification. Once the mass
becomes larger than $2H^2$, the spin-2 perturbations 
do not contain a ghost. In fact for $Hr_c =1$, 
once the second brane is put and it cuts off 
the bulk spacetime, the mass becomes $m^2 >2H^2$
regardless of the distance between the two 
branes. Note that even in the presence 
of the second brane, the Friedman equation on the 
visible brane is unchanged and the brane can 
self-accelerate. 

However, as soon as the mass of the discrete mode
of the spin-2 perturbations exceeds $m^2 = 2 H^2$, 
it is proven that the spin-0 perturbation becomes 
a ghost \cite{Izu}. At the critical length between the 
two branes where the mass of the discrete mode of 
the spin-2 perturbations is given by $m^2 =2H^2$, the 
spin-0 perturbations mix with the helicity-0 
component of the spin-2 perturbations exactly in the 
same way as the self-accelerating universe in one 
brane model and there appears a ghost. 

The spin-0 perturbation is the radion that describes 
the distance between two branes. Then one would 
try to remove this ghost by eliminating the 
radion by stabilizing the distance between 
two branes. Once we stabilize the radion, 
the brane fluctuation mode with mass $m^2=2H^2$  
becomes non-physical.
The simplest way to achieve the stabilization 
is to introduce a scalar field in the bulk. 
Then instead of the brane fluctuation, 
there appears scalar field perturbations
which can have a discrete mode and an infinite 
ladder of massive modes. In general, the mass of the 
discrete mode 
is different from $-4H^2$ which corresponds to 
$m^2=2H^2$ in the spin-2 perturbations language. 
However, it turns out that if the mass of the discrete 
mode of the spin-2 perturbations becomes $m^2 = 2H^2$, 
the discrete mode of the scalar field perturbations has 
a mass $m^2 = -4H^2$ which is the special case where
it can mix with the spin-2 perturbations. 
Then completely the same phenomena happens as in the 
one brane model. 
Once the mass of the discrete mode of spin-2 perturbations
exceeds $2H^2$, the helicity-0 mode of the spin-2 
perturbation becomes non-ghost but the scalar field 
perturbation becomes a ghost! Then it is impossible 
to remove the ghost. In the next section, we will 
explain why it is impossible to remove 
the spin-2 ghost and spin-0 ghost simultaneously. 

\section{Spectrum with matter source}
In the previous section, we identify the 
origin of the ghost by the analysis of the 
spectrum without matter source. The most 
interesting finding is that it is impossible 
to remove the spin-2 ghost and the spin-0 
ghost simultaneously \cite{Cha, Izu}. We can clarify the 
reason for this by studying the spectrum 
with matter source. We also derive the 
effective theory for perturbations on 
small scales. 

\subsection{Amplitude}
We introduce matter source on the brane 
\cite{Cha, Def5, Izu}. 
The junction condition for the transverse-traceless modes 
including matter perturbations is given by Eq.~(\ref{junction}).
Using the equation of motion for the brane bending Eq.~(\ref{bending}), 
Eq.~(\ref{junction}) is given by
\be
\left(\partial_y - 2 H \right) h^{(TT)}_{\mu \nu} =
 -\kappa^2 \Sigma_{\mu \nu} - r_c (\B -2H^2) 
 h^{(TT)}_{\mu \nu},
\ee
where
\be
\Sigma_{\mu \nu}=T_{\mu \nu}- \frac{1}{3} H^2\gamma_{\mu \nu} T
+ \frac{1}{3} (\nabla_{\mu} \nabla_{\nu} + H^2 \gamma_{\mu \nu})
(\B +4 H^2)^{-1} T.
\ee
The solution for transverse-traceless perturbations can be 
obtained by the Green's function which can be constructed 
by solutions without source obtained in section II. 
The solution is written as
\be
h_{\mu\nu}^{(TT)} = - 2 \kappa^2 \sum_i 
\frac{u_i(0)^2}{\B - 2H^2 -m_i^2} \Sigma_{\mu \nu}.
\label{TTsol}
\ee
Here the solution for the properly normalized mode functions $u_i(0)^2$ 
are given by
\bea
u_d^2(0) &=& \frac{1}{2 r_c} \frac{3 Hr_c -2}{3 Hr_c -1}, 
\quad m_d^2 = \frac{3 Hr_c -1}{r_c^2},
\nonumber\\
u_m^2(0) &=& \frac{H}{\pi} \frac{k^2}{\left(\frac{m^2 r_c}{H} - \frac{3}{2}\right)^2 + k^2},
\quad m^2 > \frac{9 H^2}{4}, \quad k = \sqrt{\frac{m^2}{H^2}- \frac{9}{4}}.
\eea
We should bear in mind that in the TT gauge, the brane 
is not located at $y=0$ and the brane bending must be 
taken into account. The induced metric on the brane is 
given by
\begin{equation}
h^{induced}_{\mu \nu}(0) = h^{TT}_{\mu \nu}
-2 H \gamma_{\mu \nu} \varphi.
\label{induced}
\end{equation}

The one particle exchange amplitude is defined as 
\be
{\cal A} \equiv \frac{1}{2} h^{induced}_{\mu \nu}(0) T^{\mu \nu} 
= \frac{1}{2} h_{\mu \nu}^{(TT)} T^{\mu \nu} - H \varphi T.
\ee
Using Eq.~(\ref{TTsol}) and Eq.~(\ref{bending}), ${\cal A}$ is calculated as 
\bea
{\cal A} 
&=& -\kappa^2 \sum_i u_i(0)^2 \left[    
T_{\mu \nu} \frac{1}{\B -2 H^2 - m_i^2} T^{\mu \nu} 
- \frac{1}{3} T \frac{1}{\B + 6 H^2 - m_i^2} T \right.\nonumber\\
 &&+ \left.\frac{1}{3} H^2 T \frac{1}{(\B + 6H^2 - m_i^2)(\B +4 H^2)}
   T \right] \nonumber\\
&&-\frac{H}{1-2 Hr_c} \frac{\kappa^2}{6} T \frac{1}{\B+4 H^2} T.
\label{ampDGP}
\eea

The second line becomes a double pole when $m_i^2 = 2H^2$
which occurs in the self-accelerating universe $Hr_c=1$
\cite{Def5}.
Usually, a double pole can be recast into difference of 
two simple poles, giving rise to a ghost. 
In fact, we can rewrite the second line as 
\be
T \frac{1}{(\B+6H^2-m_i^2) (\B+4H^2)} T
=  \frac{1}{m_i^2-2H^2} T
\left(\frac{1}{\B+6H^2 - m_i^2}-\frac{1}{\B+4H^2}\right) T.
\label{02}
\ee
The first term has a pole at $\B = m_i^2 -6 H^2$ which corresponds
to the spin-2 contributions. The second term has a pole 
at $\B = -4 H^2$ which corresponds to the spin-0 
perturbation.  Around $m_i^2 = 2H^2$, 
these terms become dominant in the amplitude and 
determine the existence of the ghost. 
A crucial point is that the contribution of 
the spin-0 perturbation is always {\it opposite} to that 
of the spin-2 massive perturbations \cite{Izu}. 
We have already seen that the massive spin-2 perturbations
contain the helicity-0 mode that is a ghost if $0 <m_i^2< 2 H^2$.
In fact we see that the sign of the amplitude changes 
at $m_i^2 = 2 H^2$. Then for $0 < m_i^2 < 2 H^2$, the  
spin-2 perturbation mediates the repulsive force. 
This means that spin-0 perturbation is not a ghost as it 
must mediate the {\it opposite} force compare to massive 
spin-2 perturbations. On the other hand, for $2 H^2 <m_i^2$,  
the spin-2 perturbations do not carry a ghost and 
mediate a normal force. Then 
the spin-0 perturbations should mediate a 
repulsive force and it should be the ghost. 
This explains the reason why the spin-0 ghost appears 
as soon as the spin-2 ghost disappears. 

In the self-accelerating universe, there appears 
a double pole \cite{Def5}. This manifests the fact that the 
spin-2 and spin-0 perturbations degenerate. In fact 
if we try to separate the poles for the spin-0 and 
spin-2 perturbations, we encounter a divergence. Then 
we need a careful treatment for this special case. 

It should be emphasized that the existence of spin-0
contribution is crucial to have a non-singular amplitude.
It is instructive to compare the amplitude (\ref{ampDGP}) 
with that in massive Pauli-Fierz theory \cite{Por}
\be
{\cal A}_{massive}
=-\kappa_4^2\left[   
T_{\mu \nu} \frac{1}{\B-2H^2 -M^2} T^{\mu \nu}
-\frac{1}{3} \frac{M^2-3 H^2}{M^2 -2H^2}
T \frac{1}{\B + 6H^2 - M^2} T
\right].
\ee
We can check that the spin-2 contributions in (\ref{ampDGP}) 
are the summation of these massive spin-2 perturbations.
The amplitude is singular for $M^2 = 2 H^2$. 
This comes from the fact that in Pauli-Fierz 
theory, we cannot couple gravity to matter with non-vanishing 
trace of energy momentum tensor. This is because the 
equation of motion is given by \cite{Por}
\begin{equation}
(2H^2-M^2)(\B + 4 H^2) h = \frac{8H^2 \kappa_4^2}{3} T,
\end{equation}
where $h$ is the trace of perturbations and 
$T$ must vanish for $M^2 = 2 H^2$. This is 
the origin of the singularity in the amplitude
and this is a pathology of massive gravity 
theory in de Sitter spacetime.  
In the DGP model, it is the brane bending mode that
couples to the trace of energy momentum tensor and 
there is no such a restriction even if the 
discrete mode has a mass $m_i^2 = 2H^2$. 
In fact the singularity at $m_i^2 = 2 H^2$
is exactly canceled by the spin-0 perturbations
as can be seen in Eq.~(\ref{02}). Thus in the DGP,
we do not have the pathology and we can 
couple gravity to matter perturbations with non-zero 
trace of energy-momentum tensor. However, 
we have seen that it is this mechanism that 
makes the spin-0 perturbation a ghost when 
the spin-2 perturbation becomes non-ghost: 
in order to cancel the singularity in the 
spin-2 sector at $m_i^2 = 2 H^2$, the spin-0 
contribution must be {\it opposite} to that 
of the spin-2 perturbations. 

\subsection{Effective theory on small scales}
On small scales, it is possible to derive the 
4D effective theory for perturbations. 
Let us consider the limit
\be
\B \gg H^2, m_i^2.
\label{assump}
\ee
Then the amplitude is approximated as 
\be
{\cal A}=-\kappa^2 \sum_i u_i(0)^2 
\left[ T_{\mu \nu} \frac{1}{\B} T^{\mu \nu}   
- \frac{1}{3} T \frac{1}{\B} T \right]
- \frac{\kappa^2}{6} \frac{H}{1-2Hr_c} T \frac{1}{\B} T.
\ee
Note that the self-accelerating universe $Hr_c =1$ is not a 
special case anymore. 
Now using the solutions for the mode functions we get 
\bea
\sum_i u_i(0)^2 
&=& u_d(0)^2 + \sum u_m(0)^2 \nonumber\\
&=& \frac{1}{2 r_c} \frac{3 Hr_c-2}{3 Hr_c -1} + \frac{H}{\pi (Hr_c)^2}
\int_{0}^{\infty} dk \frac{k^2}
{\left(k^2 + \frac{9}{4} \right) \left(k^2 + \frac{9}{4} -m_d^2 \right)} \nonumber\\
&=& \frac{1}{2r_c}.
\eea
The effective gravitational coupling is read as 
\be
\kappa^2 \sum_i u_i(0)^2 = \kappa_4^2.
\ee
Thus the 4D gravity is recovered by the summation of massive gravitons. 
Then the amplitude is calculated as 
\be
{\cal A} =-\kappa_4^2 \left[T_{\mu \nu} \frac{1}{\B} T^{\mu \nu}
- \frac{1}{3} T \frac{1}{\B} T \right]
-\frac{\kappa_4^2}{3} \frac{Hr_c}{1-2Hr_c} T \frac{1}{\B} T.
\label{limit}    
\ee
The first terms is exactly the same as the amplitude 
of massive gravity in the Minkowski spacetime. 
Due to the scalar polarisation, the coefficient 
in front of $T \B^{-1} T$ is $1/3$ not $1/2$. The last term represents 
the effect of the curvature of the brane. Then finally we get
\be
{\cal A}
=-\kappa_4^2 \left[ T_{\mu \nu} \frac{1}{\B} T^{\mu \nu}
- \frac{1}{3} \frac{1-3Hr_c}{1-2Hr_c} T \frac{1}{\B} T \right].
\label{ampsmall}
\ee
This result can be compared with the 4D Brans-Dicke (BD) theory. 
In the BD theory with BD parameter $\omega$, the amplitude
is given by
 \be
 {\cal A}
=-\kappa_4^2 \left[ T_{\mu \nu} \frac{1}{\B} T^{\mu \nu}
- \frac{1}{3} \frac{1+ \omega}{1 + \frac{2 \omega}{3}} T \frac{1}{\B} T \right].
\ee
Then the BD parameter is determined as 
\be
\omega = - 3Hr_c.
\ee
This agrees with the results obtained in Refs.~\cite{LS1, Lue2, KK3, KM}.
It is known that the BD theory contains a ghost if 
\be
\omega < -\frac{3}{2}.
\ee
This means that there is a ghost if $Hr_c > 1/2$, which 
agrees with the spectrum analysis.

\subsection{Boundary effective action}
The analysis in the previous section indicates that 
it is possible to derive the 4D effective action
on small scales. Let us again consider 
the limit
\be
\B \gg r_c^{-2}, \quad H^2
\ee
Using this approximation, we can only keep the 4D terms 
in the 5D second order action (\ref{5Dactionp1}) and (\ref{5Dactionp2}). 
Then the 4D boundary effective action is obtained as 
\bea
S_B &=& 
\frac{1}{\kappa^2} \int d^4 x \sqrt{-\gamma} 
\Big[  
(1-2 {\cal H} r_c) 
\left(h_{\mu \nu} \nabla^{\mu} \nabla^{\nu} \varphi - h 
\nabla^{\rho} \nabla_{\rho} \varphi -3 H^2 h \varphi \right) 
\nonumber\\
&&- 3{\cal H} \left(-(1-2 {\cal H} r_c)\varphi (\B + 4H^2) \varphi 
+ \frac{\kappa^2}{3} T \varphi \right)  \nonumber\\
&&+ \left. \frac{1}{2} \kappa^2 h^{\mu \nu} T_{\mu \nu}  
-\frac{r_c}{2} h^{\mu \nu} X_{\mu \nu}(h)
\right],
\eea
where ${\cal H} = H$ in the self-accelerating branch.
It is possible to check that this action consistently 
reproduce the amplitude (\ref{ampsmall}).  
We can diagonalise the action by defining 
\be
h_{\mu \nu} = \chi_{\mu \nu} - r_c^{-1}(1-2 {\cal H} r_c) \gamma_{\mu \nu} \varphi.
\ee
The resultant action is 
\be
S_B = \frac{1}{2 \kappa_4^2} \int d^4 x \sqrt{-\gamma} 
\Bigg[-\chi_{\mu \nu} X^{\mu \nu} (\chi)
+ \kappa_4^2 \chi_{\mu \nu} T^{\mu \nu} + \frac{3}{2 r_c^2}\left\{ 
(1-2Hr_c) \varphi (\B +4H^2) \varphi
- \frac{\kappa^2}{3} \varphi T \right\} \Bigg].
\label{Kaction}
\ee
The sign of the kinetic term $\varphi$ changes at $Hr_c = 1/2$
and we find that the brane bending mode $\varphi$ becomes a ghost 
for $Hr_c >1/2$. 

\section{Boundary effective action}
In the previous section, we show that it is possible to 
derive the 4D effective action. Refs.\cite{LPR} and 
\cite{NR} derived the 4D effective action for the brane 
bending mode including the non-linear interactions 
and identified the scale at which the linearized 
analysis cannot be trusted. 

\subsection{The effective action for the brane bending}
Refs.~\cite{LPR} and \cite{NR} derived the boundary effective 
action for perturbations on Minkowski background, 
$g_{MN} = \eta_{MN} + h_{MN}$.
They integrated out the bulk to obtain 
an effective action for the 4D field living on the 
boundary. The final result at the quadratic order 
is 
\be
S_b = \frac{1}{4 \kappa_4^2}
\left( \frac{1}{2} \chi_{\mu \nu} \B \chi^{\mu \nu} 
-\frac{1}{4} \chi \B \chi - r_c^{-1} n_{\mu} \triangle 
n^{\mu} + 3 r_c^{-2} \pi \B \pi \right),
\ee
where $\triangle$ is a non-local differential operator,
$\triangle = \sqrt{-\B}$ and the kinetic terms have been
diagonalized by defining 
\begin{equation}
h_{yy}= -2 \triangle \pi, \quad n_{\mu} = N_{\mu}
-\partial_{\mu} \pi, \quad \chi_{\mu \nu} = h_{\mu \nu}- 
r_c^{-1} \pi \eta_{\mu \nu},
\label{metricpi}
\end{equation} 
and $N_{\mu}$ is a shift which is $(y, \mu)$-component of the 5D metric, $g_{y \mu}$. 
In this gauge, $\pi$ plays the same role as the brane bending $\varphi$.
Note that this result is consistent with the 
previous effective action (\ref{Kaction}) for tensor and scalar 
parts if one takes $H=0$.
 
By taking into account bulk interaction with higher 
powers of $h_{MN}$, one finds that the leading order 
boundary interaction terms is cubic in $\pi$,
and involves four derivatives,
\be
S^{(3)} = - \frac{1}{2 \kappa^2} 
\int d^4 x (\partial \pi)^2 \B \pi.
\ee

In order to extract the non-trivial non-linear 
interactions, let us consider the case where 
the flat approximation is good $\chi_{\mu \nu} 
\ll 1$. However we want to preserve the 
self-coupling of the $\pi$ field. In terms 
of canonically normalized field $\hat{\pi}$ defined by
$\hat{\pi} = \pi/(2 \kappa_4 r_c)$,  
the cubic self-coupling is unchanged if 
we keep 
\be
\Lambda = \left( \frac{2 M_5^6}{M_4} \right)^{1/3}
= \left(\frac{M_4}{2 r_c^2}\right)^{1/3}, \quad 
\kappa^2 = \frac{1}{M_5^3}, \quad \kappa_4^2 = \frac{1}{M_4^2},
\ee
fixed (see below). 
We also preserve the interaction between $\pi$
and the energy momentum tensor $T_{\mu \nu}$. 
The interaction between $h_{\mu \nu}$ and $T_{\mu \nu}$
is given by $ (1/2) h_{\mu \nu} T^{\mu \nu}$. From the  
definition of $\pi$, $\hat{\pi}$ interacts with matter 
via $ (1/M_4) \hat{\pi} T$. Therefore, if we take 
the formal limit 
\be
M_4 \to \infty, \quad r_c \to \infty, \quad T_{\mu \nu} 
\to \infty, \quad 
\Lambda= {\mbox const}, \quad \frac{T_{\mu \nu}}{M_4}
= {\mbox const.}  
\label{decoupling}
\ee
we can decouple 4D gravity while keeping the full 
Lagrangian for $\hat{\pi}$. It is possible 
to check that all further interactions other than
the cubic interactions vanish. 
Then the full action for the $\hat{\pi}$ field 
in the flat spacetime is given by
\begin{equation}
S = \int d^4 x \left[   
-3     (\partial \hat{\pi})^2 - \frac{1}{\Lambda^3}
(\partial \hat{\pi})^2 \B \hat{\pi} + \frac{1}{M_4} 
\hat{\pi} T \right].
\end{equation}
From this action, the equation of motion for $\pi$
is derived as 
\be
3 \B \hat{\pi} - \frac{1}{\Lambda^3} (\partial_{\mu} 
\partial_{\nu} \hat{\pi})^2 + \frac{1}{\Lambda^3}
(\B \hat{\pi})^2 = -\frac{T}{2 M_4}.
\label{pieq}
\ee

It is possible to understand this equation from 
a geometric point of view. The Gauss-Codazzi
equation in the bulk is given by
\begin{equation}
{}^{(4)}R + K_{\mu \nu} K^{\mu \nu} - K^2 =0.
\label{GC}
\end{equation}
The junction condition at the brane is given by
\begin{equation}
K_{\mu \nu} - K \eta_{\mu \nu} 
= -\frac{\kappa^2}{2} (T_{\mu \nu} - \kappa_4^{-2} G_{\mu \nu}).
\label{cojun}
\end{equation}
Then combining these equations we get 
\begin{equation}
\frac{3 K}{r_c} + K_{\mu \nu} K^{\mu \nu} - K^2 = -\frac{T}{M_4^2}.
\label{master}
\end{equation}
Note that  this equation is exact as this 
is a combination of a geometric identity (\ref{GC})
and the junction condition on the brane (\ref{cojun}). 
From Eq.~(\ref{metricpi}), the extrinsic curvature is 
calculated as 
\be
K_{\mu \nu} = - \frac{1}{r_c \Lambda^3} \partial_{\mu}
\partial_{\nu} \hat{\pi},
\ee
in the limit (\ref{decoupling}).
Then Eq.~(\ref{master}) gives the equation of Eq.~(\ref{pieq}).

Once a solution of Eq.~(\ref{pieq}) or Eq.~(\ref{master}) is found, 
we can perturb $\hat{\pi}$ and expand the 
action up to quadratic order in the perturbations
$\rho$. We find 
\be
\delta S= \int d^4 x \left[
-3 (\partial \rho)^2 - 2 (\tilde{K}_{\mu \nu}
- \eta_{\mu \nu} \tilde{K}) \partial^{\mu} \rho
 \partial^{\nu} \rho \right].
\label{actionrho}
\ee
where $\tilde{K}_{\mu \nu} = r_c K_{\mu \nu}$ and 
it satisfies 
\be
3 \tilde{K} + \tilde{K}_{\mu \nu} \tilde{K}^{\mu \nu} 
-\tilde{K}^2 
= - \frac{T}{2 \Lambda^3 M_4}.
\ee

\subsection{Self-accelerating universe}
Let us consider a de Sitter background solution with tension
$\sigma$. The equation for $\tilde{K}_{\mu \nu}$ admits 
two branches of solutions:
\begin{equation}
\tilde{K}^{\mu}_{0 \;\nu} = 
{\cal H} r_c \delta^{\mu}_{\nu}, \quad {\cal H} 
 \equiv \frac{1\pm \sqrt{1+ 4 \,r_c^2 \sigma
/3 M_4^2}}{2 r_c} \delta^{\mu}_{\nu}. \label{Hpm}
\end{equation}
This corresponds to de Sitter solution where the Hubble 
parameter $H$ is given by
\begin{equation}
H = |{\cal H}|
\end{equation}
The self-accelerating branch corresponds to 
$+$ sign in Eq.~(\ref{Hpm}).
It is also possible to derive a corresponding solution for $\hat{\pi}$.
The de Sitter invariant solutions are 
\begin{equation}
\hat{\pi}_0 = - \frac{{\cal H} r_c}{2} \Lambda^3 x^\mu x_\mu.
\end{equation}

Now let us consider a small fluctuations around the 
de Sitter spacetime. The action for the fluctuations
$\rho$ (\ref{actionrho}) is given by
\be
\delta S= -\int d^4 x  \left[ 
3 \left(  
1 -2 {\cal H} r_c \right) 
(\partial \rho)^2 \right].
\ee
This agrees with the result for perturbations 
around de Sitter solution in the self-accelerating 
branch in the limit $\B \gg H^2$. In particular, 
the fluctuation $\rho$ becomes a ghost when 
$Hr_c > 1/2$ in the self-accelerating universe. 
In fact this was the first discovery 
of the ghost in the self-accelerating universe 
in literatures. Note that for large ${\cal H} r_c$, 
the kinetic term becomes large and the scalar mode
becomes non-dynamical. In the BD theory, this 
corresponds to the limits where the BD parameter 
$\omega = -3 {\cal H} r_c$ becomes infinite. Then
we recover GR at the linearized level \cite{Defcos}.
In this limit, the non-linear interactions are also 
suppressed. 

\subsection{Non-linearity interactions}
Next, we study the effect of the non-linear interactions 
of $\hat{\pi}$. Let is consider a static localized 
source of mass $M$. We look for a static spherically 
symmetric solution $\hat{\pi}(r)$ \cite{LPR, NR}. 
The equation for $\hat{\pi}$ (\ref{pieq}) is then given by 
\cite{MS, Koy3}
\begin{equation}
\left(\frac{d^2}{d r^2} + \frac{2}{r} \frac{d}{dr}
\right) (3 \beta \hat{\pi} + \Xi) = \frac{\rho}{2 M_4},
\label{pieqr}
\end{equation}
where $\beta =1-2 {\cal H} r_c$ and 
\begin{equation}
\Xi = \frac{2}{\Lambda^3} \int \frac{1}{r} 
\left( \frac{d \hat{\pi}}{d r} \right)^2 dr.
\end{equation}
Then it is possible to integrate the equation
to get 
\begin{equation}
3 \beta \hat{\pi} + \Xi + \frac{M_4}{2} \frac{r_g}{r}=0,
\label{phiinteg} 
\end{equation}
where 
\begin{equation}
r_g = \frac{1}{M_4^2} \int^r_0 dr r^2 \rho,
\end{equation}
is the Schwarzschild radius of the source. 
Hereafter, we assume $r_g=$ const, for simplicity. 
Taking the $r$ derivative of Eq.~(\ref{phiinteg}) gives an algebraic 
equation for $d \hat{\pi}/dr$. Then we get a solution for 
$d \hat{\pi}/dr$ as 
\begin{equation}
\frac{d \hat{\pi}}{dr} = 
\frac{3}{4} \beta \Lambda^3 r 
\left(\sqrt{1+ \left(\frac{r_*}{r}\right)^3} -1 \right),
\label{spherical:phi}
\end{equation}
where
\begin{equation}
r_*= \left(\frac{8 r_c^2 r_g}{9 \beta^2} \right)^{1/3},
\end{equation}
which is the Vainshtein radius for a source. 
On scales larger than $r_*$, the non-linear interactions 
can be neglected. On the other hand, on scales smaller 
than $r_*$ the non-linear interactions cannot be neglected 
and the linearized analysis cannot be trusted. Note that 
$r_*$ is much larger than $r_g$ if $r_c \sim H_0$ where 
$H_0$ is the present-day horizon scale. 

\subsection{Spherically symmetric solution}
Let us study the behaviour of the spherically 
symmetric solution \cite{Gru, LS1, Lue2, MS, Koy3}. The metric perturbations
are given by
\begin{equation}
h_{\mu \nu} = \chi_{\mu \nu} + \frac{2}{M_4} \hat{\pi} \eta_{\mu \nu},
\end{equation}
where $\chi_{\mu \nu}$ satisfies the 4D GR equations. 
We take the scalar perturbations on a brane as 
\begin{equation}
ds^2 = -(1 + 2 \Psi) dt^2 +
e^{2Ht}(1 + 2 \Phi) \delta_{ij} dx^i dx^j.
\end{equation}
Then the solutions for the metric perturbations can be obtained as 
\begin{eqnarray}
\Phi &=& \frac{r_g}{2r} + \frac{1}{M_4} \hat{\pi},\\
\Psi &=& -\frac{r_g}{2r} + \frac{1}{M_4} \hat{\pi}.
\label{spherical:metric}
\end{eqnarray}
On scales larger than the Vainshtein radius 
$r > r_*$, the solutions are given by
\begin{eqnarray}
\Phi &=& \frac{r_g}{2r} 
\left(1- \frac{1}{3 \beta} \right), \\ 
\Psi &=& -\frac{r_g}{2r}
\left(1+ \frac{1}{3 \beta} \right).
\end{eqnarray}
These solutions can be described by BD theory with BD
parameter $\omega = 3(1-\beta)/2 = - 3 {\cal H} r_c$.
This agrees with the analysis in the previous section. 
On scales smaller than the Vainshtein radius, $r < r_*$, the 
cubic interaction cannot be neglected. Then the linearized 
analysis cannot be trusted. In this case, the solutions for 
$\Psi$ and $\Phi$ are obtained as 
\begin{eqnarray}
\Phi &=& \frac{r_g}{2r} + \mbox{sign} (\beta)
\sqrt{\frac{r_g r}{2r_c^2}}, \\
\Psi &=& -\frac{r_g}{2r} + \mbox{sign} (\beta) 
\sqrt{\frac{r_g r}{2 r_c^2}}.
\label{einsteinphase}
\end{eqnarray}
In this region, the corrections to the solution in 4D GR 
are suppressed for $r < r_*$ so that Einstein gravity 
is recovered. From Eq.~(\ref{pieqr}), we can see that $\Xi$ dominates 
over the linear term in this region. This indicates 
that once $\hat{\pi}$ becomes non-linear, the solutions for 
the metric approach those in 4D GR. 
We should note that $\beta$ is negative in the self-accelerating 
solution while $\beta$ is positive in the normal branch solution.
Then the corrections to 4D GR solutions
have opposite signs in these solutions, as was first pointed 
out in Ref~\cite{LS1}. This means that the metric perturbations
on small scales $r \ll H^{-1}$ is sensitive to the cosmological
background solutions. We should note that Ref.~\cite{Gab2}
claimed that these solutions do not satisfy the full set 
of the non-linear equations. However, Ref.~\cite{Koy3} showed that 
these solution are fully consistent with the 5D non-linear 
equations as long as we consider scales larger than 
$r_g$. 

\section{Non-perturbative solutions}
In the previous section, we saw that the non-linear interaction
of the brane bending mode is important below Vainstein length 
$r_*$. This also means that as long as we consider length scales 
larger than Vainstein radius, the linearized analysis can be trusted 
and we find a ghost. However, in massive gravity theory, it is 
known that the non-linear interactions are very subtle. As in the 
DGP model, the perturbative approach that takes into account
non-linear interactions of the scalar mode (helicity-0 
excitations) shows that the solution approaches GR near 
the source \cite{Vai, massive}. However numerical attempts to find fully 
non-perturbative solutions have failed in massive gravity 
theory \cite{Dam}. Then it is required to 
study fully non-linear solutions carefully to check the 
validity of the linear perturbations \cite{Dva3}. 

\subsection{Schwarzschild solution} 
The attempt to find fully non-perturbative Schwarzschild 
solution was made in Ref.~\cite{Gab}. 
They assume that the 5D metric takes the form
\begin{equation}
ds^2 = - e^{\nu} dt^2 + e^{\lambda} dr^2 
+ r^2 d \Omega^2 +2 \gamma dy dr + e^{\sigma} dy^2,
\end{equation}
where $y$ is the extra-dimensional coordinate. 
In general, we should solve 5D Einstein equations and impose 
the junction conditions to find solutions. Instead of 
solving the bulk, they assumed $\nu = -\lambda$ 
to close the equations without solving the 
bulk equations. Then, remarkably, they can manage to 
find analytic solutions for $\lambda$. In the 
normal branch, below the Vainstein radius, 
the solution is given by \cite{Gab, Gab00}
\begin{equation}
- \lambda = - \frac{r_g}{r} 
- 0.84 \left(\frac{r}{r_c} \right)^2 
\left(\frac{r_*}{r}\right)^{2 (\sqrt{3}-1)},
\end{equation}
which is similar to the solution obtained in 
the previous section. However, on scales larger 
than Vainstein scale, they find 
\be
-\lambda = - \frac{\tilde{r}_g^2}{r^2}, \quad \tilde{r_g} \sim r_g \left(
\frac{r_c}{r_g}\right)^{1/3},
\label{GI}
\ee
which is the 5D solution and quite different from 
the previous 4D BD solution. Especially the 
mass of the source for observers located at 
$r > r_*$ is `screened'. In a naive perturbative
approach, we would expect that the large scale 
Newton potential behaves as 
\begin{equation}
-\lambda = -\frac{r_{g,5}^2}{r^2}, \quad 
r_{g,5}^2 = 2 G_5 M = 2 r_c r_g.
\end{equation}
In the solution (\ref{GI}) the 5D gravitatinal length is smaller 
$\tilde{r}_g^2 \ll r_{g,5}^2$. This indicates that 
the effective mass $\tilde{M}$ measured at $r > r_*$
is screened by $\tilde{M} = M (r_g/r_c)^{1/3}$. 

If this solution is a true non-perturbative solution
in the DGP model, this means that 
the linearized perturbations cannot be trusted 
even at $r > r_*$. However, we should bear in 
mind that the solution is obtained by closing 
the equations by an {\it ad hoc} metric ansatz. 
It is far from trivial that the solution obtained 
in this way satisfies the reasonable boundary conditions
in the bulk. In fact it is pointed out that the 
regularity condition in the bulk is crucial to 
specify the behaviour of gravity on the brane \cite{KM}. 
It is an open question to check that the solution
obtained in Ref.~\cite{Gab} has a regular bulk solution.

However, this solution also shows interesting 
features in the self-accelerating background 
that might be related to the ghost. In the 
self-accelerating universe, the effective mass
measured at $r > r_*$ becomes {\it negative}. 
This suggests that the self-accelerating background may not be 
problem-free even in the full non-linear 
theory.

\subsection{Domain wall}
The other example of the exact non-linear 
solutions is a domain wall. Ref.~\cite{Dva4} found 
an exact domain wall solution. In the 4D
spacetime, the domain wall creates a jump 
in the extrinsic curvature. On the other hand, 
the domain wall in the 5D spacetime is a co-dimension 
2 object like a string. Then the domain wall creates a 
deficit angle in 5D spacetime. The relation between
the deficit angle $\delta$ and the 
tension of the domain wall $\mu$ is given 
by
\be
2 H r_c \tan \gamma
\pm \gamma = \frac{\sigma}{4M_5^3},
\ee
where $\gamma$ is related to the deficit angle as 
$\delta = \mp 4 \gamma$, where $-$ is for the self-accelerating 
branch and $+$ is for the normal branch.  The first term 
comes from the jump of the 3D extrinsic curvature 
in the brane as in 4D GR and the second term is the 
contribution from the deficit angle in the 5D spacetime. 

For small tension $\sigma \ll M_5^3$, the deficit 
angle is given by
\begin{equation}
\delta = \frac{1}{1 -2 {\cal H} r_c} 
\frac{\sigma}{M_5^3}, \quad {\cal H} = \pm H.
\end{equation}
In the normal branch, there is a screening of 
the tension. The deficit angle in the 5D spacetime 
is smaller than expected. On the other hand, in the 
self-accelerating branch solution, there is 'over-screening'. 
This indicates that the wall in the self-accelerating universe 
behaves as with {\it negative} tension. Then it appears that the 
stability of the solution is not granted. This may 
be related to the existence of the ghost. In fact, 
we see that the factor that determines the 
screening $1 -2 {\cal H} r_c$ is exactly the 
factor that determines the existence of the ghost.

\section{Conclusions and Discussions}
In this review, we discuss the ghost problem in the 
self-accelerating universe. 
First we studied the spectrum of the perturbations 
without matter perturbations about a de Sitter brane. 
For the positive tension brane, the ghost comes from 
the fact that there is a discrete mode for 
the spin-2 perturbations with mass $0 < m^2 <2 H^2$.
The massive spin-2 perturbations contain a helicity-0 mode 
that becomes a ghost if the mass is in the range $0 < m^2 <2 H^2$. 
In the self-accelerating universe without tension,
the mass becomes $m^2 = 2H^2$. This is a special mass
in Pauli-Fierz massive gravity theory because there 
exists an `enhanced symmetry' that eliminates the 
helicity-0 mode. However, in the DGP model, there is 
a spin-0 perturbation with the same mass and 
this breaks the symmetry, leading to a ghost 
from the mixing between the spin-0 and spin-2 
perturbations. For a negative
tension brane, the spin-0 perturbation becomes a ghost 
if $Hr_c > 1/2$. It is easy to remove the 
spin-2 ghost by putting a second 
brane in the bulk and make the distance between the two 
branes small. Then the mass of the discrete 
mode of the spin-2 perturbations becomes 
larger than $2H^2$. However we find that 
the spin-0 perturbation, the radion, 
becomes a ghost. If we stabilize the radion, 
the perturbations of the scalar field 
that is necessary to stabilize the radion 
becomes a ghost. 

We recovered the same result by studying the spectrum
with matter source. The one particle exchange amplitude is the 
summation of massive spin-2 perturbations 
and a spin-0 perturbations. The amplitude 
of the massive spin-2 perturbations diverges 
if the mass is $m^2=2H^2$. In the DGP model, 
the spin-0 perturbation exactly cancels this singularity
at $Hr_c=1$ where the spin-2 mass becomes $m^2 = 2H^2$. 
This means that the spin-0 interaction
is opposite to that of massive spin-2 perturbations. 
This is the reason why the spin-0 perturbation 
becomes a ghost once the massive spin-2 perturbation
becomes healthy. At small scales, the effective 
theory for perturbations is described by the 
BD theory with BD parameter $\omega = -3 Hr_c$.
Here the BD scalar is a mix of the spin-0 
perturbation and the helicity-0 component 
of spin-2 perturbations. 
In the BD theory, the BD scalar mode becomes 
a ghost if $\omega <-3/2$ and this condition 
is given by $Hr_c > 1/2$ which agrees with the 
spectrum analysis. This scalar 
can be identified as the brane bending mode. 
It is possible to construct the effective action for the brane 
bending by keeping the boundary terms in the 
full 5D action. 

The effective action for the brane bending 
was shown to be a powerful tool to 
analyze the non-linear interactions 
of the scalar mode. In the decoupling 
limits where gravity decouples from the 
scalar interactions, the cubic interaction
is the dominant non-linear interaction. 
For a local source with the gravitational 
length $r_g$, the scalar mode becomes 
strongly coupled below the Vainstein 
length $r_* = (r_g r_c^2)^{1/3}$. 
Then the linearized perturbations 
cannot be trusted below this length scale. 
We checked that the effective action 
for the brane bending reproduces the 
results obtained in the linearized 
analysis beyond $r_*$. Especially, 
the brane bending mode becomes a 
ghost in this linear regime if $Hr_c >1/2$.

This seems to indicate that the linearized 
analysis makes sense at least at  
large scales $r>r_*$ and we cannot 
avoid the ghost. However, in massive 
gravity theory, fully non-perturbative 
effects are very subtle and it is possible 
that the linearized solution cannot be 
matched to the fully non-perturbed 
solutions. There are two known
fully non-perturbed solutions. One is 
Schwarzschild solution obtained in Ref.~\cite{Gab}.
Although it is an open question that this solution
has a physical solution in the bulk, the solution does 
show that the effective mass measured at 
$r > r_*$ becomes negative in the self-accelerating 
universe. This may be related to the ghost in the 
perturbative solution. The other solution is 
a domain wall solution. In this case, the exact 
5D solution is known. For a small domain wall 
tension, the domain wall has a negative deficit 
angle from 5D point of view in the self-accelerating 
branch solutions. This is yet another 
evidence that the self-accelerating universe 
suffers from the ghost instability even at 
non-perturbative level. 

There are many open questions that deserve further 
studies. We will address some of issues.\\
\hspace{1cm}\\
{\it Instabilities}\\

The ghost has a wrong sign for its kinetic term.
This means that the energy density can be 
indefinitely negative. This can lead the 
classical instability of the system. 
In Ref.~\cite{Cha} it is argued that the self-accelerating 
universe must be classically unstable. 
However, the ghost is found at linearized level 
and so far no classical instability was found at 
linearized level. We should carefully 
study the non-linear interactions to address the 
classical stability of the model and this is still
an open question. See Ref.~\cite{GKMP} for recent discussions
on non-linear instabilities.

Quantum mechanically, the ghost leads to spontaneous pair 
creation of ghosts and normal particles. Once such a channel opens, 
Lorentz invariance leads to a divergence of the particle creation 
rate and the decay rate of the vacuum is infinite. 
The same problem occurs in so-called phantom cosmology where 
the ghost scalar field is introduced to explain the 
dark energy with equation of state smaller than $-1$.
It is argued that we can avoid the rapid decay of the 
vacuum if there is a Lorentz non-invariant UV cut off 
of the order MeV in the phantom cosmology \cite{Cli2}. 
In the DGP model, the situation is more subtle \cite{IKPT}. 
If we consider the situation in which there is a spin-2 ghost, we
need to treat the helicity zero mode in a different way.
Otherwise, negative norm states appear. But if we take a different
prescription for the quantization for the helicity zero mode, this
procedure necessarily breaks de Sitter invariance. When there is a
spin-0 ghost, a similar phenomenon happens. In this case, the mass
of the ghost is given by $-4H^2$. But we know that there is no de
Sitter invariant vacuum state for a scalar field with negative
mass squared. Once de Sitter invariance is broken, one may be
allowed to consider the possibility that the non-covariant cutoff
scale may arise due to the strong coupling effect. The 
strong coupling length is very large $\Lambda^{-1} \sim 1000$ km.
Then the particle creation could be milder than the usual ghost 
in the Minkowski background \cite{IT}. Clearly, further studies are necessary to 
verify this.

\hspace{1cm}\\
{\it The fate of instabilities}\\

If the self-accelerating branch solutions have the ghost
instability, then we are naturally lead
to ask what does this solution decay to. 
An interesting fact is that the normal branch solutions are 
ghost-free. Then it is
tempting to think that the self-accelerating solutions decay into
the normal branch solutions. In fact for a given 
tension, the Hubble parameter in the self-accelerating 
universe is larger than that of the normal branch solutions.
Then there could be the nucleation
of bubbles of the normal branch in the environment of the
self-accelerating branch solution. This would resemble a kind of false
vacuum decay in de Sitter space. False vacuum decay is described
by an instanton which is a classical solution in an Euclidean time
connecting initial and final configurations. In our case, we are
interested in a solution that interpolates between the
self-accelerating and the normal branches. 
Ref.~\cite{IKPT} tried to construct such an instanton solution. It was 
found that the solution requires the presence of a 2-brane (the
bubble wall) which induces the transition.
However, this instanton cannot be realized as the
thin wall limit of any smooth solution. Once the bubble thickness
is resolved, the equations of motion do not allow $O(4)$ symmetric
solutions joining the two branches.
It was concluded that the thin wall instanton is unphysical, and that
one cannot have processes connecting the two branches. 
This suggests that the self-accelerating branch does not decay into the
normal branch by forming normal branch bubbles. 
Thus it is still unclear what is the end state of the ghost instability. 

\hspace{1cm}\\
{\it Possible ways out of the ghost?}\\

In Ref.~\cite{Def5}, several ways out of the ghost problem were 
discussed. In the self-accelerating universe
$Hr_c =1$, there could be the enhanced symmetry that 
can eliminate the helicity-0 ghost if there were no 
spin-0 perturbation. Ref.~\cite{Def5} performed 
a gauge transformation 
\begin{equation}
A_{\mu \nu} = B_{\mu \nu} +(\nabla_{\mu} \nabla_{\nu} + H^2 \gamma_{\mu \nu})
X, \quad X= \frac{3}{4H} \varphi.
\end{equation}
Then the action becomes 
\be
\begin{split}
S_{eff}=\frac{1}{\kappa^2 H}\int d^4x\sqrt{-\gamma}
\bigg\{&-B^{\mu\nu}X_{\mu\nu}(B)-H^2B^{\mu\nu}B_{\mu\nu}
+H^2B^2\\
&- \varphi H (\nabla_{\mu} \nabla_{\nu}-\gamma_{\mu \nu} \B
-3 H^2 \gamma_{\mu \nu} ) B^{\mu \nu} \bigg\}.
\end{split}
\ee
They propose to treat the spin-0 perturbation $\varphi$ 
as a Lagrangian multiplier and perform quantization 
a la Nakanishi-Lautrup way in the QED. Effectively, this 
removes $\varphi$ from the spectrum and the enhanced symmetry 
ensures that there is no helicity-0 mode and no ghost 
(see Ref.~\cite{Pad} for a criticism on this procedure). 
However, this method cannot be applied if there is 
matter fluctuations with non-zero trace of energy-
momentum tensor. This is because the trace of 
energy-momentum tensor breaks the enhanced symmetry 
and excite $\varphi$ as we saw in section IV.A. 

The other way to remove the ghost is to take into
account the non-normalizable modes in the spectrum \cite{Def5}.
Then it is possible to eliminate 
the double pole from the amplitude that was the origin of 
the ghost. Instead, the amplitude has a single pole 
corresponding to the non-normalizable massless mode. 
However, it is found that this 
massless mode has an opposite sign for the 
amplitude compared with continuous massive states. 
This indicates that this massless mode becomes
ghost-like. A very similar phenomena was found in 
the analysis of the shock wave analysis where 
the non-normalizable modes contribute a repulsive 
potential \cite{Kal, Cha}. 

\hspace{1cm}\\
{\it Modifying the model}\\

Finally, in order to remove the ghost from the 
theory, we may have to modify the starting action.
There are several attempts to extend the model
\cite{Izu, Car, Rha, Gabnew, Kos}. 
As we explained in section III.D, it is impossible 
to remove the ghost in two-brane model \cite{Izu}. 
It was also shown that the introduction of Gauss-Bonnet 
term in the bulk does not help \cite{Rha}.
Recently, a new model was proposed where the bulk 
is the solution in the normal branch but the 
junction condition is the one in the self-accelerating 
branch \cite{Gabnew}. In order to achieve this, the 5D Einstein
-Hilbert action with a opposite sign to the conventional 
one is assumed to be localized near the brane. It remains to be seen 
whether this model can evade the ghost or not by studying 
the perturbations. The other approach would be 
consider higher co-dimensional branes \cite{higher} or 
intersecting branes \cite{CKT}. It still remains 
to be seen whether there exists the self-accelerating 
universe and there is no ghost in these models. 
It is also important to study whether it is possible to embed the DGP model 
in string theory \cite{string}. A UV completion of the model is necessary 
to study the fate of the ghost instability and it can guide us to make 
some simple modification of the model that will ameliorate the problems 
discussed in this review \cite{GKMP}.


\begin{thebibliography}{}
\bibitem{SN}
   B.~P.~Schmidt {\it et al.}  [Supernova Search Team Collaboration],
  Astrophys.\ J.\  {\bf 507}, 46 (1998)
  [arXiv:astro-ph/9805200].
  A.~G.~Riess {\it et al.}  [Supernova Search Team Collaboration],
  Astron.\ J.\  {\bf 116}, 1009 (1998)
  [arXiv:astro-ph/9805201];
  S.~Perlmutter {\it et al.}  [Supernova Cosmology Project Collaboration],
  Astrophys.\ J.\  {\bf 517} 565 (1999) 
  [arXiv:astro-ph/9812133].
    
\bibitem{SN2}
  P.~Ruiz-Lapuente,
  Class. Quantum. Grav. {\bf 24} R91 (2007)
  [arXiv:0704.1058 [gr-qc]].

\bibitem{CMB}
  D.~N.~Spergel {\it et al.}  [WMAP Collaboration],
  arXiv:astro-ph/0603449.

\bibitem{LSS}
  M.~Tegmark {\it et al.}  [SDSS Collaboration],
  Phys.\ Rev.\  D {\bf 69}, 103501 (2004)
  [arXiv:astro-ph/0310723].
  
\bibitem{RM}
  R.~Maartens,
  arXiv:astro-ph/0602415.

\bibitem{KK0}
  K.~Koyama,
  arXiv:0706.1557 [astro-ph].

\bibitem{Dva}
  G.~R.~Dvali, G.~Gabadadze and M.~Porrati,
  Phys.\ Lett.\  B {\bf 485}, 208 (2000)
  [arXiv:hep-th/0005016].

\bibitem{Aka0}
  K.~Akama,
  Prog.\ Theor.\ Phys.\  {\bf 60}, 1900 (1978).

\bibitem{Deffayet:2000uy}
  C.~Deffayet,
  Phys.\ Lett.\ B {\bf 502}, 199 (2001)
  [arXiv:hep-th/0010186].

\bibitem{Def2}
  C.~Deffayet, G.~R.~Dvali and G.~Gabadadze,
  Phys.\ Rev.\  D {\bf 65} (2002) 044023
  [arXiv:astro-ph/0105068].

\bibitem{Deffayet:2002sp}
  C.~Deffayet, S.~J.~Landau, J.~Raux, M.~Zaldarriaga and P.~Astier,
  Phys.\ Rev.\ D {\bf 66}, 024019 (2002)
  [arXiv:astro-ph/0201164].

\bibitem{Def3}
  C.~Deffayet, G.~R.~Dvali, G.~Gabadadze and A.~I.~Vainshtein,
  Phys.\ Rev.\ D {\bf 65} (2002) 044026
  [arXiv:hep-th/0106001];

\bibitem{Lue}
  A.~Lue,
  Phys.\ Rev.\ D {\bf 66} (2002) 043509
  [arXiv:hep-th/0111168].

\bibitem{Gru}
  A.~Gruzinov,
  New Astron.\  {\bf 10} (2005) 311
  [arXiv:astro-ph/0112246];

\bibitem{Tan}
  T.~Tanaka,
  Phys.\ Rev.\ D {\bf 69} (2004) 024001
  [arXiv:gr-qc/0305031].

\bibitem{LS1}
 A.~Lue and G.~Starkman,
  Phys.\ Rev.\ D {\bf 67} (2003) 064002
  [arXiv:astro-ph/0212083].
  
\bibitem{Lue2}
  A.~Lue, R.~Scoccimarro and G.~D.~Starkman,
  Phys.\ Rev.\ D {\bf 69}, 124015 (2004)
  [arXiv:astro-ph/0401515].

\bibitem{MS}
  C.~Middleton and G.~Siopsis,
  Mod.\ Phys.\ Lett.\ A {\bf 19}, 2259 (2004)
  [arXiv:hep-th/0311070].


\bibitem{Koy3}
  K.~Koyama and F.~P.~Silva,
  Phys.\ Rev.\  D {\bf 75}, 084040 (2007)
  [arXiv:hep-th/0702169].

\bibitem{LueR}
  A.~Lue,
  Phys.\ Rept.\  {\bf 423}, 1 (2006)
  [arXiv:astro-ph/0510068].

\bibitem{Rubakov}
  V.~A.~Rubakov,
  arXiv:hep-th/0303125;

\bibitem{LPR}
  M.~A.~Luty, M.~Porrati and R.~Rattazzi,
  JHEP {\bf 0309} (2003) 029
  [arXiv:hep-th/0303116];

\bibitem{NR}
  A.~Nicolis and R.~Rattazzi,
  JHEP {\bf 0406} (2004) 059
  [arXiv:hep-th/0404159];

\bibitem{KK1}
  K.~Koyama,
  Phys.\ Rev.\ D {\bf 72} (2005) 123511
  [arXiv:hep-th/0503191].

\bibitem{KK2}
  D.~Gorbunov, K.~Koyama and S.~Sibiryakov,
  Phys.\ Rev.\ D {\bf 73} (2006) 044016
  [arXiv:hep-th/0512097].

\bibitem{Cha}
  C.~Charmousis, R.~Gregory, N.~Kaloper and A.~Padilla,
  JHEP {\bf 0610} (2006) 066
  [arXiv:hep-th/0604086].

\bibitem{Def5}
  C.~Deffayet, G.~Gabadadze and A.~Iglesias,
  JCAP {\bf 0608} (2006) 012
  [arXiv:hep-th/0607099].

\bibitem{Izu}
  K.~Izumi, K.~Koyama and T.~Tanaka,
  [arXiv:hep-th/0610282].


\bibitem{KK-0}
  K.~Koyama and K.~Koyama,
  Phys.\ Rev.\ D {\bf 72} (2005) 043511
  [arXiv:hep-th/0501232].

\bibitem{Tanaka}
J.~Garriga and T.~Tanaka, Phys. Rev. Lett. {\bf 84} (2000) 2778; 
T.~Tanaka, Phys. Rev. {\bf D69} (2004) 024001.

\bibitem{Padilla}
A.~Padilla,
Class.\ Quant.\ Grav.\  {\bf 21} (2004) 2899.

\bibitem{PF}
M.~Fierz and W.~Pauli, Proc. Roy. Soc. {\bf 173} (1939) 211.

\bibitem{vDVZ}
H.~van Dam and M.~Veltman, Nucl. Phys. {\bf B22} (1970) 397; 
V.~I.~Zakharov, JETP Lett. {\bf 12} (1970) 312.

\bibitem{Higuchi}
A.~Higuchi, Nucl. Phys. {\bf B282} (1987) 397; 
S.~Deser and R.~I.~Nepomechie, Ann. Phys. {\bf 154} (1984) 396. 

\bibitem{DW}
S.~Deser and A~Waldron, Phys. Lett. {\bf B508} (2001) 347.

\bibitem{Por}
  M.~Porrati,
  Phys.\ Lett.\  B {\bf 498} (2001) 92
  [arXiv:hep-th/0011152].

\bibitem{KK3}
  K.~Koyama and R.~Maartens,
  JCAP {\bf 0601} (2006) 016
  [arXiv:astro-ph/0511634].

\bibitem{KM}
  K.~Koyama and S.~Mizuno,
  JCAP {\bf 0607}, 013 (2006)
  [arXiv:gr-qc/0606056].

\bibitem{Defcos}
  C.~Deffayet,
  Phys.\ Rev.\  D {\bf 71}, 023520 (2005)
  [arXiv:hep-th/0409302].

\bibitem{Gab2}
  G.~Gabadadze and A.~Iglesias,
  Phys.\ Lett.\  B {\bf 639}, 88 (2006)
  [arXiv:hep-th/0603199].

\bibitem{Vai}
A.~I.~Vainshtein, Phys.\ Lett.\ B {\bf 39}, 393 (1972).

\bibitem{massive}
  N.~Arkani-Hamed, H.~Georgi and M.~D.~Schwartz,
  Annals Phys.\  {\bf 305}, 96 (2003)
  [arXiv:hep-th/0210184].

\bibitem{Dam}
  T.~Damour, I.~I.~Kogan and A.~Papazoglou,
  Phys.\ Rev.\  D {\bf 67}, 064009 (2003)
  [arXiv:hep-th/0212155].

\bibitem{Dva3}
  G.~Dvali,
  New J.\ Phys.\  {\bf 8} (2006) 326
  [arXiv:hep-th/0610013].

\bibitem{Gab}  
  G.~Gabadadze and A.~Iglesias,
  Phys.\ Rev.\ D {\bf 72} (2005) 084024
  [arXiv:hep-th/0407049].

\bibitem{Gab00}
  G.~Gabadadze and A.~Iglesias,
  Phys.\ Lett.\  B {\bf 639}, 88 (2006)
  [arXiv:hep-th/0603199].

\bibitem{Dva4}
  G.~Dvali, G.~Gabadadze, O.~Pujolas and R.~Rahman,
  [arXiv:hep-th/0612016].

\bibitem{GKMP}
  R.~Gregory, N.~Kaloper, R.~C.~Myers and A.~Padilla,
  arXiv:0707.2666 [hep-th].
  
 \bibitem{Cli2}
  J.~M.~Cline, S.~Jeon and G.~D.~Moore,
  Phys.\ Rev.\  D {\bf 70}, 043543 (2004)
  [arXiv:hep-ph/0311312].

\bibitem{IKPT}
  K.~Izumi, K.~Koyama, O.~Pujolas and T.~Tanaka,
  arXiv:0706.1980 [hep-th].

\bibitem{IT}
K.~Izumi and T.~Tanaka, arXiv:0709.0199 [gr-qc].

\bibitem{Pad}
  A.~Padilla,
  arXiv:hep-th/0610093.

\bibitem{Kal}
  N.~Kaloper,
  Phys.\ Rev.\ Lett.\  {\bf 94}, 181601 (2005)
  [Erratum-ibid.\  {\bf 95}, 059901 (2005)]
  [arXiv:hep-th/0501028];
  N.~Kaloper,
  Phys.\ Rev.\  D {\bf 71}, 086003 (2005)
  [Erratum-ibid.\  D {\bf 71}, 129905 (2005)]
  [arXiv:hep-th/0502035].

\bibitem{Rha}
  C.~de Rham and A.~J.~Tolley,
  JCAP {\bf 0607}, 004 (2006)
  [arXiv:hep-th/0605122].

\bibitem{Car}
  M.~Carena, J.~Lykken, M.~Park and J.~Santiago,
  Phys.\ Rev.\  D {\bf 75}, 026009 (2007)
  [arXiv:hep-th/0611157].

\bibitem{Gabnew}
  G.~Gabadadze,
  arXiv:hep-th/0612213;
  G.~Gabadadze,
  arXiv:0705.1929 [hep-th].

\bibitem{Kos}
A.~S.~Koshelev and T.~N.~Tomaras,
arXiv:0706.3393 [hep-th].


\bibitem{higher}
  G.~R.~Dvali and G.~Gabadadze,
  Phys.\ Rev.\  D {\bf 63}, 065007 (2001)
  [arXiv:hep-th/0008054];
  S.~L.~Dubovsky and V.~A.~Rubakov,
  Phys.\ Rev.\  D {\bf 67}, 104014 (2003)
  [arXiv:hep-th/0212222];
  M.~Kolanovic, M.~Porrati and J.~W.~Rombouts,
  Phys.\ Rev.\  D {\bf 68}, 064018 (2003)
  [arXiv:hep-th/0304148];
  G.~Gabadadze and M.~Shifman,
  Phys.\ Rev.\  D {\bf 69}, 124032 (2004)
  [arXiv:hep-th/0312289];
  N.~Kaloper and D.~Kiley,
  arXiv:hep-th/0703190.

\bibitem{CKT}
O.~Corradini, K.~Koyama and G.~Tasinato, to appear.

\bibitem{string}
  E.~Kiritsis, N.~Tetradis and T.~N.~Tomaras,
  JHEP {\bf 0108}, 012 (2001)
  [arXiv:hep-th/0106050];
  I.~Antoniadis, R.~Minasian and P.~Vanhove,
  Nucl.\ Phys.\  B {\bf 648}, 69 (2003)
  [arXiv:hep-th/0209030].

\end{thebibliography}
\end{document}